\documentclass[longbibliography,aps,preprint,onecolumn,letterpaper,superscriptaddress,amsmath,amsfonts,showpacs,floatfix]{revtex4}
\usepackage{graphicx}
\usepackage{amsmath, esint}
\usepackage{color, hyperref, natbib}
\usepackage{lipsum}

\renewcommand{\vec}{\mathbf }
\newcommand{\R}{\mathbb R}
\newcommand{\C}{\mathbb C}
\newcommand{\e}{\eqref}

\newcommand{\diver}{\mathrm{div}}

\newcommand{\D}{\mathrm d}
\newcommand{\I}{\mathrm i}
\newcommand{\pD}{\partial }

\DeclareMathOperator{\atan}{atan}

\begin{document}

\title{Superharmonic Instability of Stokes Waves}

\date{\today}

\author{A.\,O.~Korotkevich}
\email{alexkor@math.unm.edu}
\affiliation{Department of Mathematics and Statistics, University of New Mexico, MSC01 1115, 1 University of New Mexico, Albuquerque, NM 87131-0001, USA}
\affiliation{L.\,D.~Landau Institute for Theoretical Physics RAS, Prosp. Akademika Semenova 1A, Chernogolovka, Moscow region, 142432, Russian Federation}

\author{P.\,M.~Lushnikov}
\email{plushnik@math.unm.edu}
\affiliation{Department of Mathematics and Statistics, University of New Mexico, MSC01 1115, 1 University of New Mexico, Albuquerque, NM 87131-0001, USA}

\author{A.~Semenova}
\email{asemenov@uw.edu}
\affiliation{ICERM, Brown University, Box E 11th Floor, 121 South Main Street, Providence, RI 02903, USA}

\author{S.\,A.~Dyachenko}
\email{sergeydy@buffalo.edu}
\affiliation{Department of Mathematics, University at Buffalo (SUNY), 244 Mathematics Building, Buffalo, NY 14260-2900, USA}


\begin{abstract}
A  stability of  nearly limiting Stokes waves to superharmonic
perturbations  is considered numerically. The new, previously inaccessible
branches of superharmonic instability were investigated. Our numerical simulations
suggest that eigenvalues of linearized dynamical equations, corresponding
to the unstable modes, appear as a result of a collision of a pair of purely imaginary
eigenvalues at the origin, and a subsequent appearance of a pair of purely real eigenvalues:
a positive and a negative one that are symmetric with respect to zero. Complex conjugate
pairs of purely imaginary eigenvalues correspond to stable modes, and as the steepness of the
underlying Stokes wave grows, the pairs move toward the origin along the imaginary axis.
Moreover, when studying the eigenvalues of linearized dynamical we find that as the steepness of
the Stokes wave grows, the real eigenvalues follow a universal scaling law, that can
be approximated by a power law. The asymptotic power law behaviour of
this dependence for instability of Stokes waves close to the limiting one
is proposed. Surface elevation profiles for several unstable eigenmodes
are made available through~\url{http://stokeswave.org} website.
\end{abstract}


\maketitle

\section{Introduction}
A key object in ocean dynamics is a swell which is a spatially periodic train of surface
gravity waves propagating in one direction with constant velocity. Such train is well described by Stokes waves discovered by Stokes~\cite{Stokes1847, Stokes1880_197,Stokes1880_314}. These nonlinear long waves (with wavelengths typically in the range from meters to hundred of meters)  propagate without change of form and are described  by a potential flow of ideal (incompressible and inviscid) $2$-dimensional fluid with free surface and infinite depth.
There is a long history of study of Stokes waves including~\cite{michell1893,Nekrasov1921,MalcolmGrantJFM1973LimitingStokes,SchwartzJFM1974,Longuet-HigginsFoxJFM1977,Longuet-HigginsFoxJFM1978,toland1978existence,Plotnikov1982,AmickFraenkelTolandActaMath1982,Williams1981,WilliamsBook1985,TanveerProcRoySoc1991,Longuet-HigginsWaveMotion2008,DyachenkoLushnikovKorotkevichJETPLett2014,DLK2016,LushnikovStokesParIIJFM2016,LushnikovDyachenkoSilantyevProcRoySocA2017}.

A stability of Stokes waves determines  an eventual fate of Stokes waves in
ocean.  We follow
\cite{Longuet-HigginsFoxProcRoySoc1978superharmonics} and
\cite{Longuet-HigginsFoxProcRoySoc1978subharmonics} to distinguish {\it\
superharmonic} stability and {\it subharmonic}  stability.  Superharmonic
stability means addressing perturbations with the same spatial period as
the spatial period $\Lambda$ of Stokes wave (with the cases of the smaller
spatial periods $\Lambda/n, n=2,3,\ldots$ of perturbations also included as particular cases).
Subharmonic perturbations have larger period than $\Lambda.$  Subharmonic
instability of  deep water with small amplitudes has been extensively
studied since~\cite{benjamin1967disintegration,Lighthill1965,whitham1967non,Zakharov1968}. The same instability was
discovered in  \cite{BespalovTalanovJETPLett1966} for nonlinear optics. That  instability is now
called either by Benjamin-Feir instability or modulational instability,
see also  \cite{ZakharovOstrovskyPhysD2009} for historical overview.
Modulational instability is efficiently described
by the approximation of nonlinear Schr\"odinger equation for the envelope
of slowly modulated Stokes wave \cite{Zakharov1968}.   A nonlinear
stage of the development of that instability results in formation of
solitons as well as in weak turbulence of surface gravity waves with
dynamics of time scales greatly exceeding a period of Stokes wave.
Another type is high frequency instability of Stokes waves of
small amplitude which typically produces small growth rates, see
 \cite{DeconinckOliverasJFM2011,creedon2022high}.
Ref.~\cite{murashige2020stability} provides a conformal mapping approach
to linear stability of Stokes waves in irrotational and waves on shear current
setting for both super- and subharmonic instability.

In this work, we  focus on superharmonic instability of strongly nonlinear Stokes waves. Instability growth rate is much larger than the growth rate of modulational instability of small amplitudes waves.  Thus this instability may play significant role in wavebreaking at the nonlinear stage of instability development. This is consistent with well-know oceanic observations, water tank experiments and large scale simulations that strongly nonlinear gravity waves quickly results in multiple wavebreaking events provided steepness $H/\Lambda$ exceeds $\approx  0.0178$ ~\cite{BT1998,BBY2000,SB2002,KPZ2019}.

A nonlinearity of Stokes wave is determined by a steepness  $H/\Lambda$,
where $H$  is the Stokes wave height defined as the vertical distance from
the crest to the trough of Stokes wave. Without loss of generality we use
scaled units  at which a phase speed $c_0$ of linear gravity wave of
wavelength $\Lambda$ is $c_0=1$  and we set  $\Lambda=2\pi$ (see
e.g.~\cite{DLK2016} for details of that scaling). In these units Stokes
wave has a speed  $c>1$ with the limit  $H\to 0, \ c\to 1$ corresponding
to the  linear gravity wave.  The Stokes wave of the greatest height
$H=H_{max}$ (also called by the limiting Stokes wave) has the singularity
in the form  of the sharp angle of $2\pi/3$ radians on the
crest~\cite{Stokes1880_314}.
Refs.~\cite{DLK2016,LushnikovDyachenkoSilantyevProcRoySocA2017} and a
website~\url{http://stokeswave.org} provide high precision numerical
approximation for Stokes wave including the estimate
$H_{max}/\Lambda=0.141063483980 \pm 10^{-12}$.

In~\cite{Longuet-HigginsFoxProcRoySoc1978superharmonics} the superharmonic instability of Stokes waves was predicted
at the steepness exceeding $H/\Lambda \approx 0.1388$  (Ref.~\cite{Longuet-HigginsFoxProcRoySoc1978superharmonics} used $ka$ for waves steepness with $k=2\pi/\Lambda$ and $a=H/2$ implying  $ka=\pi H/\Lambda$) and suggested that the instability threshold corresponds to the maximum of $c$ as the function of $H/\Lambda$.
In~\cite{tanaka1983stability} the first computation of a growth rate of superharmonic instability was performed  from the analysis of the eigenvalue problem of the linearization around Stokes wave   and found that superharmonic instability has a threshold at $H/\Lambda = 0.1366$,
with one unstable mode appearing above that threshold.
In addition, in~\cite{tanaka1983stability} it was conjectured that this threshold corresponds to the  first maximum of the total energy of Stokes wave as the function of $H/\Lambda$ in the contrast with the prediction of~\cite{Longuet-HigginsFoxProcRoySoc1978superharmonics}. This conjecture was confirmed analytically in~\cite{SaffmanJFM1985superharmonic} based on the Hamiltonian formulation of free surface dynamics~\cite{Zakharov1968}, see also  \cite{BridgesJFM2004superharmonic} for more discussion.  In~\cite{LonguetHigginsTanakaJFM1997} it was found that as steepness of the Stokes wave
increases past $H/\Lambda \approx 0.1366$ a second unstable mode appears at  $H/\Lambda \approx 0.141$.
It is natural to assume that as we approach the limiting Stokes waves, more unstable modes would appear.
A nonlinear stage of the development of  Stokes waves instability of all these modes typically results in wave breaking as  were studied from simulations  in multiple papers including~\cite{longuet1978deformation,LonguetHigginsDommermuthJFM1997, dyachenko2016whitecapping}.

In this paper we  provide numerical solution of eigenvalue problem for superharmonic instability  and obtain  three unstable branches.  These branches originate from extrema of Stokes wave energy as a function of $H/\Lambda$.
In particular, the first instability branch originates at $H_1/\Lambda=0.1366035\pm 10^{-7}$, the second branch at  $H_2/\Lambda= 0.1407965\pm 10^{-7}$ and the third branch at $H_3/\Lambda = 0.1410496\pm 10^{-7}$.
The accuracy of these numerical values can be further improved to any desired level by computing Stoke wave with variable precision following approaches of  \cite{DyachenkoLushnikovKorotkevichJETPLett2014,DLK2016,LushnikovStokesParIIJFM2016,LushnikovDyachenkoSilantyevProcRoySocA2017}. The latter paper discusses implementation of an auxiliary conformal mapping to improve the convergence rate of Fourier series representing Stokes solutions; this mapping was used to improve the numerical resolution of the eigenfunctions appearing in the linear stability analysis implemented in the present paper.
 We found that the dependence of these growth rates as functions of $H/\Lambda$  collapses into\ a universal curve via a shift and a rescaling of $H/\Lambda$    into  $(H_{max} - H)/(H_{max}-H_n)$, where $n = 1,2,3$ is the number of the unstable eigenmode.

The paper is organized as follows. Section \ref{sec:dynamicalequations}
provides basic dynamic equations for free surface dynamics. Section
\ref{sec:linearization}  {considers a linearization of these equations and
formulate an eigenvalue  for the stability analysis of  Stokes wave.
Section \ref{sec:numericaleigenvalue} describes a numerical approach to
solve the large scale eigenvalue problem. A shift-invert
method in combination with Arnoldi algorithm is used to address
eigenvalues for large matrices up to $90,000\times 90,000$ of linearized
problem.  Section \ref{sec:MainResults}
provides the main results  including  the numerical results on three
unstable branches in Section \ref{sec:eigenvalues} and a rescaling
of these branches to the universal curve in Section \ref{sec:universal}. Section \ref{sec:DiscussionConclusions} summarizes the main
results and discusses future directions.

\section{Formulation of the Problem}

\subsection{Dynamical equations of free surface dynamics}
\label{sec:dynamicalequations}

We consider a 2D flow of an ideal incompressible fluid with a free surface in a gravity field without surface tension. Fluid occupies the region $-\infty<x<\infty$ and $-\infty<y<\eta(x,t)$
with the elevation of the moving free surface given
by the function $y = \eta(x,t)$ at a moment of time $t$.
A gravity field is pointed in the negative direction of $y. $ We consider a potential flow with the velocity $\vec v$ represented through
the scalar velocity potential $\Phi(x,y;t)$ as follows $\vec v = \vec\nabla\Phi$.
The incompressibility condition $\diver(\vec v)=0$ requires the velocity potential to be a harmonic
function $\nabla^2\Phi=0$.
The kinematic boundary condition (BC)  \begin{equation}
\frac{\pD \eta}{\pD t} =
\left. \left(-\frac{\pD \Phi}{\pD x} \frac{\pD \eta}{\pD x} +\frac{\partial \Phi}{\partial y} \right)\right|_{y = \eta(x,t)} \label{kinematic_dynamic_bc}
\end{equation}
and dynamic BC\begin{equation}
 \quad\;\;
\left. \left(\frac{\pD \Phi}{\pD t} +\frac{1}{2} \left( \vec\nabla \Phi \right )^2 \right)
\right|_{y = \eta(x,t)}+g\eta = 0, \label{kinematic_dynamic_bcb}
\end{equation}
have to be satisfied on the free surface, where $g$ is the acceleration due to gravity. The kinematic BC means that  the free surface moves together with fluid particles located at that surface, i.e. there is no separation of fluid particles from the free surface. The dynamic BC is given by the time-dependent Bernoulli equation at the free surface ensuring the zero pressure at the free surface.  We also assume decaying boundary condition on the velocity potential deep inside fluid
$\left.\Phi(x,y)\right|_{y\to-\infty}\to0$. We consider periodic solutions thus focusing of one period of length $\Lambda$ with $x\in[-\Lambda/2, \Lambda/2]$ and periodic
BC in $x$.

As a result, we have to solve Laplace equation $\nabla^2\Phi=0$ in
time-dependent domain with the motion of of free surface determined by boundary conditions~\eqref{kinematic_dynamic_bc},\eqref{kinematic_dynamic_bcb} which form a closed set of equation. An efficient  way to
solve these  equation is through the time-dependent conformal mapping $z(w,t)=x(u,v;t) + \I y(x,y;t)$    of a fixed domain (lower complex half-plane $\C^-$) of the auxiliary variable $w=u+\I v,$ $u,v \in \R$ into a time-dependent fluid domain in the physical complex plane $z=x+\I y$.  Because of assumed $\Lambda$-periodicity in $x$, we restrict to one spatial period (along $u$) in $w$-plane as well, see
Figure~\ref{conformal_map}.

\begin{figure}\label{conformal_map}
  \includegraphics[width=0.9\textwidth]{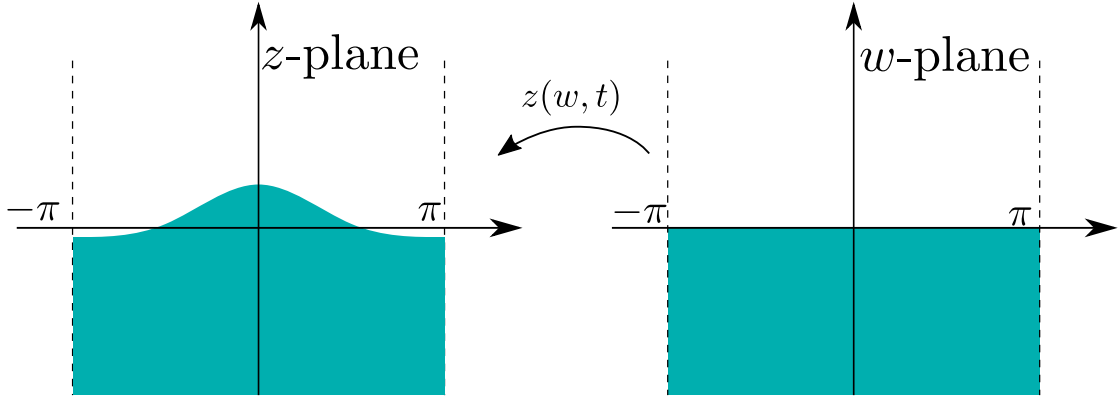}
  \caption{Half-strip in $w$ plane ($(u,v) \in [-\pi,\pi]\times(-\infty,0]$) into the area in $(x,y)$
plane under the free-surface $\eta(x,t)$.
The line $v = 0$ is mapped into the fluid surface.}
\end{figure}

The idea of using such type of time-dependent conformal transformation
   was
exploited by several authors
including~\cite{Ovsyannikov1973,MeisonOrzagIzraelyJCompPhys1981,TanveerProcRoySoc1991,TanveerProcRoySoc1993,DKSZ1996,ChalikovSheininAdvFluidMech1998,ChalikovSheininJCompPhys2005}.
We follow~\cite{DKSZ1996} to recast equations \eqref{kinematic_dynamic_bc} and \eqref{kinematic_dynamic_bcb}  into
the equivalent form for $x(u,0;t), \ y(u,0;t)$ and $\Psi(u,0;t)$ (here and below we abuse the notation and use the  same symbol $\Psi$ for function of both $(x,y,t)$ and $(u,v,t)$)
at the
real line $w=u$ of the complex plane $w$ as follows
\begin{equation} \label{ypsit-explicit}
y_t=(y_u\hat H-x_u)\left [\frac{\hat H\Psi_u}{|z_u|^2}\right ],\;\;\Psi_t=\Psi_u\hat H\left [\frac{\hat H\Psi_u}{|z_u|^2}\right ]+\frac{\hat H\left [\Psi_u\hat H \Psi_u  \right ]}{|z_u|^2}-gy,\;x=u-\hat H y,
\end{equation}
Here subscripts denote partial derivatives and  $\hat Hf(u)=\frac{1}{\pi} \text{p.v.}
\int^{+\infty}_{-\infty}\frac{f(u')}{u'-u}\D u'$ is the Hilbert transform with $\text{p.v.}$ meaning a Cauchy principal value of the integral.
The Hilbert transform
in Fourier space is given by $(\hat H f(u))_k=\I\,\text{sign}{\,(k)}\,f_k$, with $f_k$ being the harmonics of Fourier series
specified for $\Lambda$-periodic function $f(u)=f(u+\Lambda)$ as follows%
\begin{equation} \label{ffourier}
f_k=\frac{1}{\Lambda}\int\limits_{-\Lambda/2}^{\Lambda/2}
f(u)\exp\left (-\I ku\frac{2\pi}{\Lambda}\right )\D  u,\;\; f(u)=\sum\limits_{k=-\infty}^{\infty} f_k\exp\left (\I
ku\frac{2\pi}{\Lambda}\right ),
\end{equation}
where $\text{sign}(k)=-1,0,1$ for $k<0, \ k=0$ and $k>0$, respectively.

More compact  but equivalent form  of equations  \e{ypsit-explicit} was found in  ~\cite{Dyachenko2001} as follows \begin{align}
\frac{\pD R}{\pD t} &= \I \left(U R_u - R U_u \right), \label{Reqn4}\\
\qquad U &=\hat P^-(R\bar V+\bar RV), \quad B= \hat P^-(|V|^2), \label{UBdef4}\\
\frac{\pD V}{\pD t} &= \I \left[ U V_u - RB_u\right ]+ g(R-1), \label{Veqn4}
\end{align}
and is often called ``Tanveer--Dyachenko equations''.
Here the new unknowns%
\begin{equation}\label{?}
R \equiv \frac{1}{z_u}\quad\mbox{and}\quad V \equiv \frac{\I(\Psi+\I\hat H\Psi)_u}{z_u}
\end{equation}
 were introduced by S. Tanveer in
~\cite{TanveerProcRoySoc1991} for the periodic BC and later independently
obtained by A. I. Dyachenko in
\cite{Dyachenko2001} for the decaying  BCs  so we refer to these variables as ``Tanveer-Dyachenko
variables". Also $\hat P^-\equiv \frac{1}{2}(1+\I \hat H)$ is the projector operator of any function $f(u)$ defined by the Fourier series \e{ffourier} into the space of functions analytic in lower half plane $w\in \C^-$, which is given by
$\hat P^-(f(u))=f_0/2+\sum\limits_{k=-\infty1}^{-1} f_k\exp\left (\I
ku\frac{2\pi}{\Lambda}\right )$. Here and below $\bar f$ means a complex conjugate of $f.$ Equations ~\eqref{Reqn4}-\eqref{Veqn4} are convenient to consider below in a problem of stability of the
Stokes waves.

\subsection{Linearization and eigenvalue problem}
\label{sec:linearization}

Stokes wave is time-independent solution of equations  ~\eqref{Reqn4}-\eqref{Veqn4}  in the moving frame with the speed $c$ such that both $R$\ and $V$ are functions of  $u-ct$ only. To study stability of Stokes waves, we first consider a small
perturbation of general solutions  $R$, $V$ of
equations~\eqref{UBdef4}-\eqref{Veqn4} in the following form $R\rightarrow
R+\delta R$, $V\rightarrow V+\delta V$. A linearization of Eqs. \eqref{Reqn4}-\eqref{Veqn4} with respect to perturbations $\delta R$ and
$\delta V$ gives
that\begin{align}
\frac{\partial \delta R}{\partial t} &= \I \left(\delta U R_u +U \delta R_u - \delta R U_u - R \delta U_u \right), \label{Reqn4lin}\\
\qquad \delta U&=\hat P^-(\delta R\bar V+R\delta\bar V+ {\delta \bar R}V+\bar R\delta V), \quad \delta B= \hat P^-(\delta V \bar V+V\delta\bar V), \label{UBdef4lin}\\
\frac{\partial \delta V}{\partial t} &= \I \left[ \delta U V_u + U \delta
V_u - \delta RB_u- R\delta B_u\right ]+ g\delta R. \label{Veqn4lin}
\end{align}
Now we add a restriction that both $R$ and $V$ in equations \e{Reqn4lin}-\e{Veqn4lin} correspond to Stokes wave.  Assuming an exponential time dependence of perturbation around Stokes wave, we represent these perturbations as follows
\begin{equation}\label{dRdVlambda}\begin{split}
& \delta R(u-ct,t)=e^{\lambda t}\delta R_{1}(u-ct)+e^{\bar\lambda t}\delta R_{2}(u-ct), \\
& \delta V(u-ct,t)=e^{\lambda t}\delta V_{1}(u-ct)+e^{\bar\lambda t}\delta
V_{2}(u-ct),
\end{split}
\end{equation}
 where subscripts $1$ and and its complex conjugate $2$ are used to distinguish different functions of $u.$  $Re(\lambda)$ is the growth
rate of perturbation. Then
\begin{equation}\label{dRdVlambdabar}
\begin{split}
& \delta \bar R(u-ct,t)=e^{\bar \lambda t}\delta \bar R_{1}(u-ct)+e^{\lambda t}\delta \bar R_{2}(u-ct), \\
& \delta \bar V(u-ct,t)=e^{\bar \lambda t}\delta \bar
V_{1}(u-ct)+e^{\lambda t}\delta \bar V_{2}(u-ct).
\end{split}
\end{equation}
A dynamics of general perturbations can be represented as superposition of solutions with different $\lambda$. Thus our goal is to find possible values of $\lambda$.

Substituting~\eqref{dRdVlambda} and~\eqref{dRdVlambdabar}
into~\eqref{Reqn4lin}-\eqref{Veqn4lin} and collecting terms $\propto e^{\lambda t}$ we obtain that\begin{align}
\lambda \delta R_{1} &=c(\delta R_{1})_u+ \I \left[\delta U_{1}  R_u +U (\delta R_{1})_u - \delta R_{1} U_u - R( \delta U_{1})_u \right],\nonumber\\
\lambda \delta \bar R_{2} &=c(\delta \bar R_{2})_u- \I \left[\delta \bar U_{2}  \bar R_u +\bar U (\delta \bar R_{2})_u - \delta \bar R_{2} \bar U_u - \bar R( \delta \bar U_{2})_u \right], \label{RVeqn4lina}\\
\lambda \delta V_{1} &=c(\delta V_{1})_u+ \I \left[ \delta U_{1} V_u + U
(\delta
V_{1})_u - \delta R_{1}B_u- R(\delta B_{1})_u\right ]+ g\delta R_{1}, \nonumber\\
\lambda \delta \bar V_{2} &= c(\delta \bar V_{2})_u-\I \left[ \delta\bar
U_{2} \bar V_u + \bar U (\delta\bar V_{2})_u - \delta \bar R_{2}\bar B_u-
\bar R(\delta \bar B_{2})_u\right ]+ g\delta \bar R_{2},\nonumber
\end{align}
where
\begin{align}
\delta U_{1}&=\hat P^-(\delta R_{1}\bar V+R\delta\bar V_{2}+ {\delta \bar R_{2}}V+\bar R\delta V_{1}), \nonumber\\
\delta \bar U_{2}&=\hat P^+(\delta \bar R_{2} V+\bar R\delta V_{1}+
{\delta  R_{1}}\bar V+R\delta \bar V_{2}),
\nonumber \\
\quad \delta B_{1}&= \hat P^-(\delta V _{1}\bar V+V\delta\bar V_{2}), \nonumber   \\
\quad \delta\bar B_{2}&= \hat P^+(\delta\bar  V _{2} V+\bar V\delta
V_{1}).\nonumber
\end{align}
Here $\hat P^{+}(f(u))\equiv\frac{1}{2}(1-\I \hat H)f$ is the projector onto the class of functions analytic in the upper half-plane $\C^{+}$ of $w$.

Equations~\eqref{RVeqn4lina} together with  the periodicity of $\delta
R_{1}, \delta \bar R_{2}, \delta V_{1}, \delta \bar V_{2}$  in $u$ form
the eigenvalue problem for the eigenvector
\begin{equation} \label{deltaRVeigen}
(\delta R_{1}, \delta \bar R_{2}, \delta V_{1}, \delta \bar V_{2})^{T},
\end{equation}
where $T$ means transposition.
Without loss of generality we assume the spatial period  $2\pi$.

\subsection{Numerical solution of the eigenvalue problem}
\label{sec:numericaleigenvalue}

For $R$\ and $V$ in equations \eqref{RVeqn4lina} 
we use high
precision Stokes waves available at~\cite{PadePolesList}.
Eigenvalue problem given by the equations~\eqref{RVeqn4lina} were solved by application of
shift-invert method in combination with Arnoldi algorithm for largest magnitude eigenvalues,
specifically ARPACK-NG (available at~\cite{ARPACK-NG}) realization was used. We briefly describe that algorithm below.

We represent each of $\delta R_{1}, \delta \bar R_{2}, \delta V_{1},
\delta \bar V_{2}$ by a truncated Fourier series of $N$ Fourier harmonics.
Then equations~\eqref{RVeqn4lina} can be written in a matrix form as
follows:
\begin{align}
        \hat A {\vec x} = \lambda {\vec x},
\end{align}
where $\hat A$ is a $4 \times 4$ block operator matrix. It can be reduced to a matrix of coefficients $A$
by acting on the natural basis $(\vec e_i)_j = \delta_{i,j}$ in wavenumbers space with where $\delta_{i,j}$ being the Kr\"oneker delta,  $\delta_{i,j} =1$ for $i=j$ and $\delta_{i,j}=0$ for $i\ne j$.

Arnoldi algorithm is the most efficient, when it tries to locate few eigenvalues of largest magnitude. Let us suppose that we
have a guess of an eigenvalue $\sigma$. Then we can consider the modified eigenvalue problem~\cite{saad1992numerical}:
\begin{equation}
(A-\sigma I)^{-1}\vec x = \nu \vec x,\label{Arnoldi_modified}
\end{equation}
eigenvalues of which $\nu_j$ are related to the eigenvalues of original problem $\lambda_j$ by a simple formula:
\begin{equation}
\nu_j = \frac{1}{\lambda_j-\sigma}.
\end{equation}
It is clear, that if our guess $\sigma$ is close enough to the eigenvalue $\lambda_j$ we are looking for,
the magnitude of the $\nu_j$ eigenvalue will be the largest one.
In practice, it was enough to take $\sigma=0.1$ and to request to find 16 largest magnitude eigenvalues of modified problem~\eqref{Arnoldi_modified}
to find all purely real value $\lambda_j$'s corresponding to unstable eigenmodes. Instead of computation of $(A-\sigma I)^{-1}$
with multiplication on $\vec x$ it is more efficient to perform once $LU$-factorization of $A-\sigma I$ and then solve
a linear system $(A-\sigma I) {\vec v} = {\vec x}$ in order to find $\vec v = (A-\sigma I)^{-1}\vec x$. In order to decrease memory
requirements, we use our knowledge of analytic structure of parts of~\eqref{deltaRVeigen}. Specifically, in Fourier space,
all functions without complex
conjugation signs have to be analytic in the lower half plane, meaning that we can neglect harmonics with positive $k$ (pay attention, that $k=0$ harmonic
has to be kept!), while for functions with bars it is enough to keep only harmonics with positive $k$ (these functions are analytic in the upper half
of complex plane). Such approach allows to decrease the memory requirements for storage of $A$ by a factor of 4. In addition, one can also consider
iterative methods for solving of $(A-\sigma I) {\vec v} = {\vec x}$ using one of the standard algorithms for non-symmetric matrices (e.g. GMRES~\cite{SaadSchultzGMRES1986}), as application of $\hat A$ operator can be performed using $O(N\log N)$-operations, where N is the number of harmonics.
Another approach which accelerated our computations dramatically is auxiliary conformal mapping, effectively introducing nonhomogeneous grid,
which was originally introduced in~\cite{LushnikovDyachenkoSilantyevProcRoySocA2017} and is described in Appendix~\ref{atantanmap}.
Calculation with the same level of accuracy on the homogeneous grid would would require e.g. $N\sim 10^{9}$ harmonics instead of $N=45000$.

During the computations, spurious egenvalues were observed close to the origing of the complex plane with both real and imaginary parts of the order of $10^{-8}$ and smaller. One could detect them by changing the number of used harmonics, as they were slightly changing in position, while the physically relevant eigenvalues (both real and imaginary ones) were practically stationary.
Also, computations of egenvalues with different resolutions and methods allowed us to determine  how many digits of precision after decimal point we could trust (usually at least 6).

We were able to compute eigenvalues for matrices of the size up to $90,000\times 90,000$ (corresponding to resolution of $N=45000$ harmonics for the original Stokes' wave), which for complex double precision numbers corresponds to $\simeq 120$ GiB. Computations in such a case were taking more than 24 hours on a relatively modern 24-cores computational workstation and used practically all available $128$ GiB of RAM.
The memory usage could be substantially decreased by application of iterative methods
for solution of \eqref{Arnoldi_modified} instead of formation of the full matrix, as it is described above. This is a necessary improvement for investigation of next instability branches and will be done in the near future.

\section{Main Results}
\label{sec:MainResults}

For the Stokes wave it is traditional to introduce a wave
steepness $s$ as a ratio $s=H/\Lambda$ of crest-to-trough height $H$ and
the wavelength $\Lambda$. It is well-known that integral quantities associated
with the Stokes wave oscillate as a function of wave steepness $s$.
Following the asymptotic theory
of~\cite{longuet1978theory}-\cite{longuet1997crestpart3}, we may identify
the extremal points of the Hamiltonian,
\begin{align}
   E = \frac{1}{2}\int \psi\hat k \psi \, du + \frac{g}{2} \int y^2x_u \,du, \label{Heq}
\end{align}
as Stokes waves approach the wave of the greatest height.

This theory provides formulae for Stokes wave speed, and total energy, $E$, in
the vicinity of limiting wave:
\begin{align}
        &c^2(\epsilon) = \frac{g}{k}\left(1.1931 - 1.18\epsilon^3\cos(2.143\ln\epsilon + 2.22 \right), \\
        &E(\epsilon) = \frac{g}{k}\left(0.07286 - 0.383\epsilon^3\cos(2.143\ln\epsilon + 1.59 \right), \label{lhE}
\end{align}
where $\epsilon^2 = \frac{kq^2}{2g}$ provides a distinct parameterization of the Stokes wave family.
Here $k=2\pi/\Lambda$ and $q$ is the magnitude of velocity of a fluid particle located at the crest of the
wave measured in the reference frame moving with the speed $c$.
Local extrema of Hamiltonian can be obtained from formula~\eqref{lhE} as
\begin{align}
        \frac{\partial E}{\partial \epsilon} = 0, \quad\mbox{when} \quad \tan\left(2.143\ln\epsilon + 1.59\right) = 1.4.
\end{align}
\begin{table*}
        \centering
        \begin{tabular}{lccccc}
                \hline
                $n$ & $\frac{H}{\Lambda}$ \small{(Longuet-Higgins)} & $\frac{H}{\Lambda}$ \small{(numerics)} & $E$ \small{(Longuet-Higgins)} & $E$ \small{(numerics)} \\
                \hline
                1&0.136258683901074 & 0.13660355596621762 & 0.464823018228553 & 0.46517718027280353\\
                2&0.140827871097976 & 0.14079658408852538 & 0.457706391816943 & 0.45770579203963280\\
                3&0.141061656416396 & 0.14104962672339530 & 0.457793945537506 & 0.45779727678745985\\
                4&0.141074235010001 & 0.14106274069928185 & 0.457792868390338 & 0.45779615273931995
        \end{tabular}
    \caption{Correspondence between Hamiltonian extrema positions from Longuet-Higgins theory~\eqref{lhE} and numerical computations improves with increase of steepness.
    We computed the extrema of Hamiltonian and corresponding steepnesses by the $6$th order polynomial interpolation with error being of the order $10^{-7}$.}
        \label{tab:LHtheory}
\end{table*}

In the Table~\ref{tab:LHtheory} we show the comparison of the results obtained from Longuet--Higgins asymptotic theory and the results of numerical
simulations of the fully nonlinear equations for the Stokes wave. In the first column, we show the locations for extrema of the Hamiltonian at which
unstable eigenmodes occur as estimated from Longuet--Higgins formula~\eqref{lhE}, and the second column are the locations of extrema of Hamiltonian
obtained from our direct computations. The $3$rd and the $4$th columns show the corresponding values of Hamiltonian at these extrema.

\begin{figure}
        \includegraphics[width=0.495\textwidth]{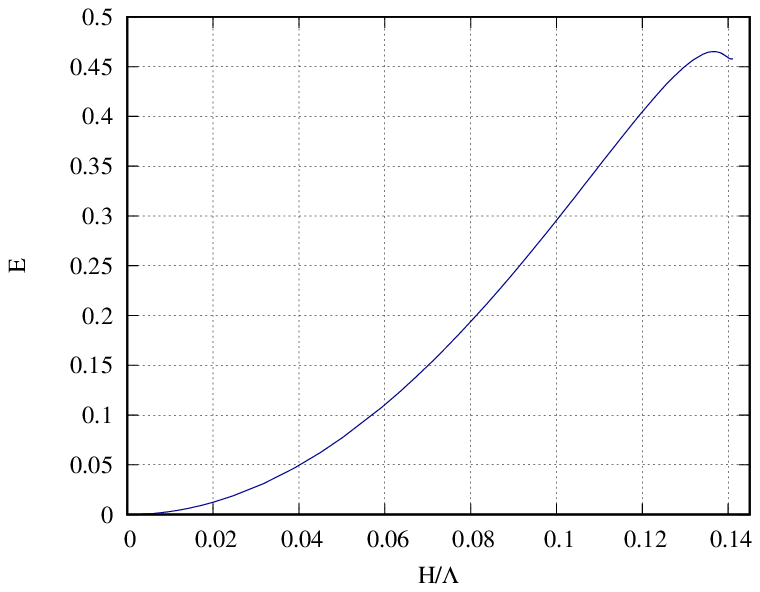}
        \includegraphics[width=0.495\textwidth]{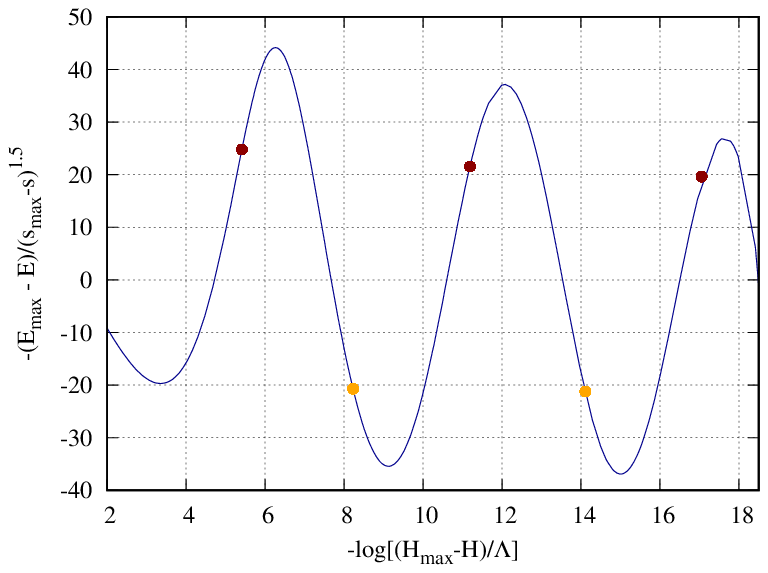}
    \caption{(Left Panel) The Hamiltonian, $E$, as a function of $s$, the steepness of the Stokes wave, exhibits oscillations in the
    vicinity of the limiting wave. (Right Panel) In order to distinguish the tiny features of oscillations we show a magnified
    plot in the logarithmic scale $-\log(s_{max} - s)$ and the Hamiltonian is magnified by a factor $(s_{max}-s)^{3/2}$. The extrema of
    the Hamiltonian are displaced to positions marked by the red points (the maxima) and the golden points (the minima).
    }
        \label{fig:Energy}
\end{figure}

It was shown in~\cite{tanaka1983stability}-\cite{longuet1997crest} that as
we increase steepness of the Stokes wave, after some threshold, there
appears the first unstable eigenmode. It was investigated in details in
the papers mentioned before. Also it was demonstrated that with further
increase of stepness the second unstable mode appears. The values of
steepness, which are thresholds for new unstable modes appearance,
correspond to the local extrema of the Hamiltonian of the Stokes wave. We
were able to investigate in details the first three unstable modes.

It is convenient to consider square of the eigenvalues corresponding to unstable modes as a function of steepness. Before the threshold eigenvalues are
purely imaginary and above the threshold they are purely real. So we can define the threshold as a point where square of the eigenvalue goes through zero.
Corresponding functions for the first two unstable modes are represented in Figure~\ref{fig:eig1}.

\begin{figure}
        \includegraphics[width=0.495\textwidth]{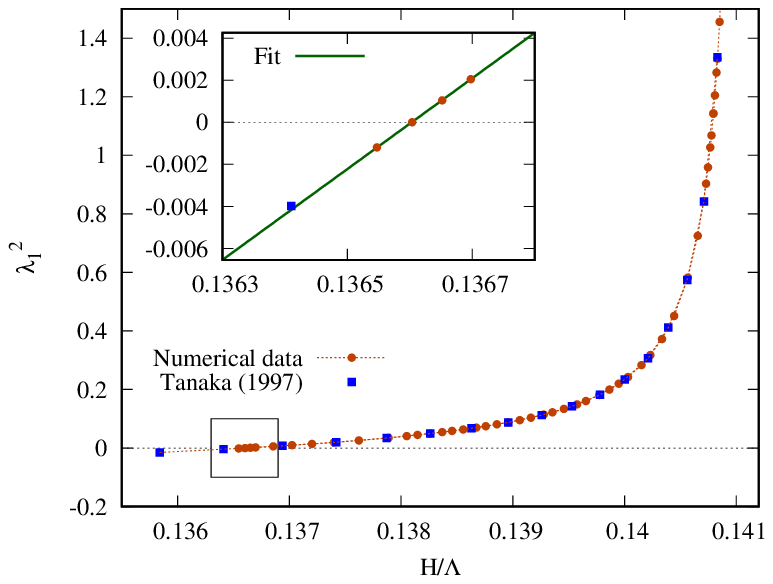}
        \includegraphics[width=0.495\textwidth]{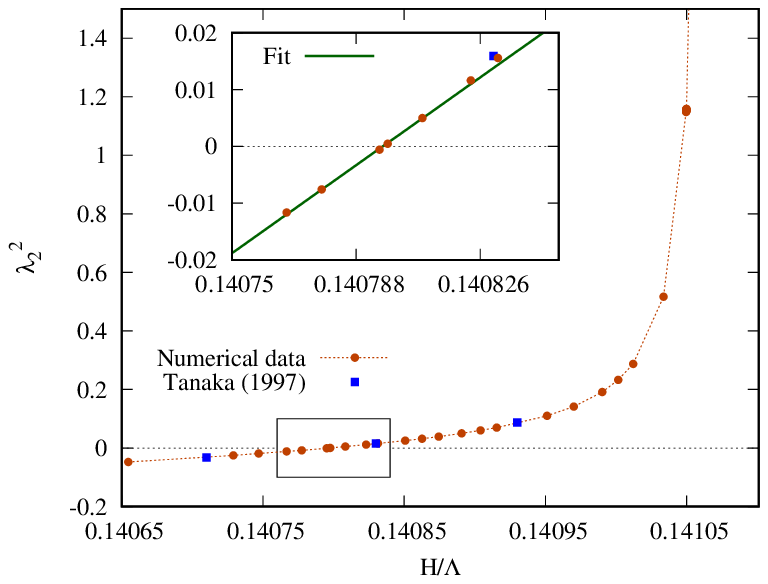}
        \caption{(Left) The square of first real eigenvalue $\lambda^2(s)$ is crossing the instability threshold at $s_1 = 0.1366035$ when Hamiltonian
        goes through the first local extremum. The eigenvalues computed in present work (orange circles), the numerical data of the work~\cite{longuet1997crest}
        (blue squares) all fit well with the same line (green line); (Right) The square of the second real eigenvalue $\lambda^2(s)$ crosses
        the instability threshold at $s_2 = 0.140796$.}
        \label{fig:eig1}
\end{figure}

We used a least square fit to a linear function:
\begin{equation}
f_n (s) \sim (s-s_n)
\label{eig:fit}
\end{equation}
in the vicinity of appearance of every eigenmode. As a result of this
procedure we were able to find thresholds for the appearance of the first
unstable mode $s_1 = 0.136603552635709$ and the second mode $s_2 =
0.140796170578837$. These numbers correspond to the values obtained from
direct Stokes waves calculations (see Table~\ref{tab:LHtheory} up to 7 and
6 digits, respectively (close to accuracy of the obtained eigenvalues). For the third unstable eigenmode the fitting
procedure gave $s_3 = 0.141049633798808$, 7 digits of which coincide with
the result of direct computations in Table~\ref{tab:LHtheory}. The plot of
squared eigenvalues is given in the left panel of Figure~\ref{fig:eig2}.

\begin{figure}
        \includegraphics[width=0.495\textwidth]{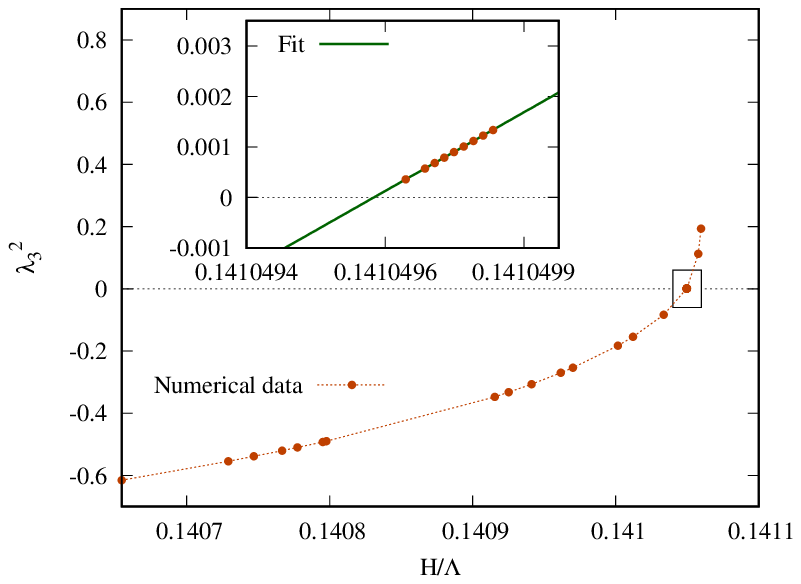}
        \includegraphics[width=0.495\textwidth]{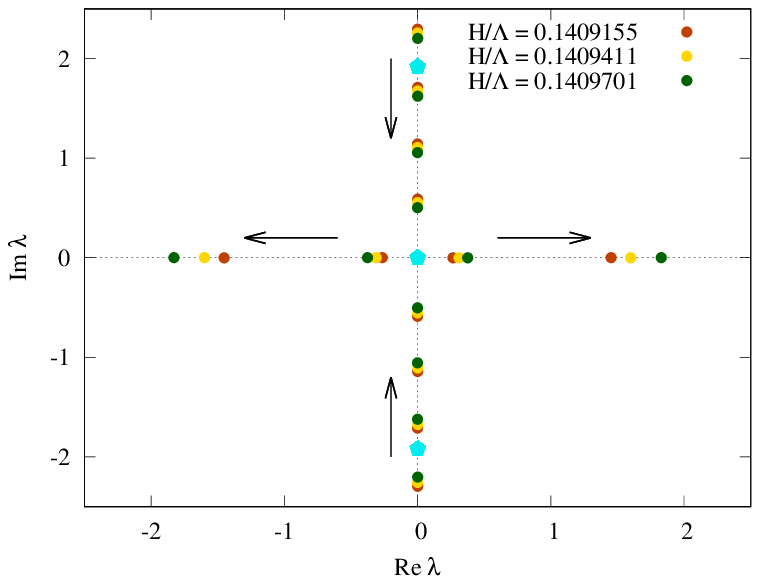}
        \caption{(Left) The square of the third real eigenvalue is crossing the instability threshold at $s_3 = 1.410496$ at the third local extremum of the
        Hamiltonian. Circles are numerical solutions of eigenvalue problem, and solid line is a numerical fit.
        (Right) A motion of eigenvalues near the origin just after the second extremum of the Hamiltonian,
        shows that there are two kinds of eigenvalues, the ones that are sensitive to small changes in $s$ (red, yellow and green), and the
        less sensitive ones (cyan). It is evident that more eigenvalues are moving to the origin to collide and produce more unstable
        eigenmodes.}
        \label{fig:eig2}
\end{figure}

\subsection{Dependence of eigenvalues on the steepness and appearance of new branches of instability}
\label{sec:eigenvalues}

In the right panel of Figure~\ref{fig:eig2} one can observe that eigenvalues continuously move in the complex plane as steepness
grows. We find that some eigenvalues are more sensitive than others to small changes of steepness of the underlying Stokes wave.
The less sensitive eigenvalues are marked with cyan pentagons, they are located at the origin and on the imaginary axis.

The green, yellow and red circles correspond to the sensitive eigenvalues that are observed as complex conjugated pairs. These
eigenvalues continuously move toward the origin as the steepness of the underlying Stokes wave grows. The first pair collides at
the origin when the steepness of the underlying Stokes wave reaches the first maximum of the Hamiltonian, the second pair collides
when the steepness of Stokes wave reaches the first minimum of the Hamiltonian, and so forth.
The mechanism of collisions with formation of unstable modes was previously discussed in the work~\cite{mackay1986stability}.

The linear dispersion relation of the gravity wave in the frame moving with velocity $c$ is given by $\omega_k=\pm ck \pm \sqrt{g k}$. It provides  a good estimate to the eigenvalues for linearization about small amplitude Stokes
waves, but we find that finite amplitude Stokes waves always have small deviations from the linear dispersion relation due to the nonlinear
frequency shift \cite{ZLF1992}. The discrepancy between the linear dispersion and less sensitive eigenvalues obtained numerically become more
evident as the steepness of the Stokes waves, hence nonlinearity of the system, grows.

\subsection{Universal dependence of branches of instability}
\label{sec:universal}

It is striking to note, that all computed eigenvalues for all eigemodes
($n = 1,2,3$) 
collapse into one curve (see Figure~\ref{fig:eig3}, left panel).
Here we used
the normalized variable $(s_{max} - s)/(s_{max}-s_n)$ on horizontal axis where
$n = 1,2,3$ is the number of the unstable eigenmode and $s_{max}$
corresponds to the steepness of the limiting Stokes wave (e.g.
see~\cite{DLK2016}).

The curve is fitted by the nonlinear least squares algorithm to the function $(b_0 + b_1x + b_2x^2 + b_3 x^3)\log(x)$ with
$b_0 = -0.140023$, $b_1 = 0.0366936$, $b_2 = -0.0129251$, and $b_3 = 0.00125835$.
In the right panel of the Figure~\ref{fig:eig3}, we
see that the eigenvalues can be well approximated by the power law $\lambda_n^2
\propto 1/(s_{max}-s)$ in the vicinity of the limiting Stokes wave for all
$n$.

\begin{figure}
        \includegraphics[width=0.495\textwidth]{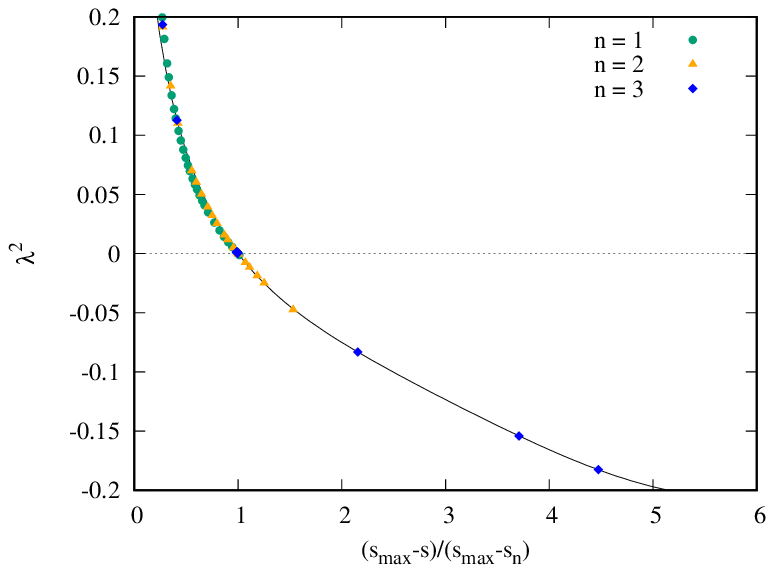}
        \includegraphics[width=0.495\textwidth]{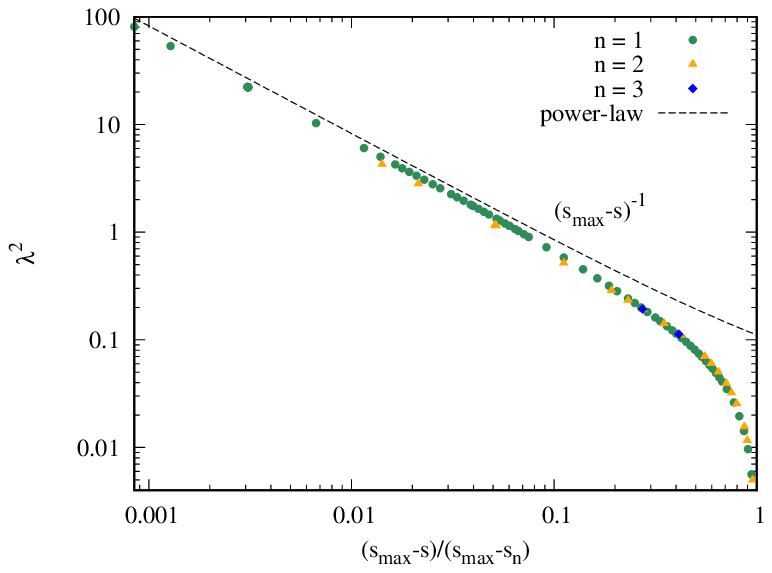}
        \caption{(Left panel) All eigenvalues corresponding to unstable eigenmodes collapse to a single universal curve as a function of normalized steepness
        $(s_{max} - s)/(s_{max}-s_n)$, where $n = 1,2,3$ is the number of the unstable eigenmode and $s_{max}$ corresponds to the steepness of the limiting Stokes wave.
        (Right panel) Plot in loglog scale of eigenvalues and power law $\lambda_n^2 \sim 1/(s_{max}-s)$ in the vicinity of the
        limiting wave for all $n$.}
        \label{fig:eig3}
\end{figure}

\section{Discussion and Conclusions}
\label{sec:DiscussionConclusions}

We compute the first three unstable eigenmodes of linearized equations
about Stokes waves with the same spatial period as in Stokes wave (superharmonic instability). It is shown that these unstable
modes emerge at  the threshold values of steepness, which correspond to
the extrema of Hamiltonian. This fact supports and extends observations
in~\cite{longuet1997crest}.
The results of numerical computations suggest
that eigenvalues corresponding to unstable eigenmodes appear due to a
collision of a pair of purely imaginary eigenvalues at the origin in the
complex plane when steepness reaches the threshold values.

Our conjecture based on the results in the Figure~\ref{fig:eig3} is
that all eigenvalues corresponding to unstable eigenmodes above and below the thresholds of instability
lie on a single curve if we plot them as a function of normalized steepness $(s_{max} - s)/(s_{max}-s_n)$.
In addition, our simulations suggest power law for $\lambda_n^2 \sim 1/(s_{max}-s)$ in the vicinity of the
limiting wave for unstable eigenmodes $n = 1, 2, 3$. Further analytical work is needed to explain these observations.

Investigation of the stability properties can give one an insight into the
evolution of the Stokes wave. In practice trains of Stokes waves in turbulent ocean
is always
accompanied by  other small waves which can be considered as perturbation of
the solution. If we can represent such perturbation as a combination of
unstable eigenmodes, the initial stage of the dynamics of the perturbation
will be determined by growth rates of the unstable eigenmodes.

The approach to investigation of instabilities of solutions similar to
Stokes waves developed in this paper can be applied to study stability of
Stokes waves with constant vorticity~\cite{dosaev2017simulation, murashige2020stability}.
We plan to extend the stability analysis of the present paper to waves
on a linear shear current formulated in conformal variables such see
e.g.~\cite{dyachenko2019stokes}.

\section*{Acknowledgments.}
The authors gratefully wish to acknowledge the following contributions:
The work of SD was supported by the National Science Foundation under Grant No. DMS-2039071. The work of P.M.L. was supported by the National Science Foundation, grant no. DMS-1814619. The AS material is based upon work supported by the National Science Foundation under Grant No. DMS-1929284 while the AS was in residence at the Institute for Computational and Experimental Research in Mathematics in Providence, RI, during the "Hamiltonian Methods in Dispersive and Wave Evolution Equations" program. Initial work of AK and SD on the subject was supported by the National Science Foundation Grant No. OCE-1131791.

Also authors would like to thank developers of FFTW~\cite{FFTW}, ARPACK-NG~\cite{ARPACK-NG}, and the whole GNU project~\cite{GNU} for developing, and supporting this useful and free software.

\appendix

\section{\label{atantanmap}Auxiliary Conformal Mapping}

We employ additional conformal mapping given by the formula:
\begin{align}
    u(q) = 2\atan{L\tan{\frac{q}{2}}},\quad\mbox{and}\quad q_u = \frac{1}{L}\left( \cos^2{\frac{q}{2}} + L^2 \sin^2{\frac{q}{2}}\right)
\end{align}
that allows to reduce the number of Fourier modes for resolving eigenfunctions of the linearization problem from $N$ to $\sim \sqrt{N}$ and
allows to find eigenvalues in the vicinity of the third extremum of the Stokes wave.

The eigenvalue problem formulated in the $q$-plane is closely related to the equations~\eqref{RVeqn4lina} and is given by:
\begin{align}
\lambda u_q \delta R_{1} &=c(\delta R_{1})_q+ \I \left[\delta U_{1}  R_q +U (\delta R_{1})_q - \delta R_{1} U_q - R( \delta U_{1})_q \right],\nonumber\\
\lambda u_q \delta \bar R_{2} &=c(\delta \bar R_{2})_q- \I \left[\delta \bar U_{2}  \bar R_q +\bar U (\delta \bar R_{2})_q - \delta \bar R_{2} \bar U_q - \bar R( \delta \bar U_{2})_q \right], \label{RVeqn4linaQ}\\
\lambda u_q \delta V_{1} &=c(\delta V_{1})_q+ \I \left[ \delta U_{1} V_q + U
(\delta
V_{1})_q - \delta R_{1}B_q- R(\delta B_{1})_q\right ]+ gu_q\delta R_{1}, \nonumber\\
\lambda u_q \delta \bar V_{2} &= c(\delta \bar V_{2})_q-\I \left[ \delta\bar
U_{2} \bar V_q + \bar U (\delta\bar V_{2})_q - \delta \bar R_{2}\bar B_q-
\bar R(\delta \bar B_{2})_q\right ]+ gu_q\delta \bar R_{2},\nonumber
\end{align}

\bibliography{StokesWave_AS,surfacewaves_KAO,surfacewaves_PL,lushnikov,biblionls,tmp}

\end{document}